\documentstyle[12pt]{article}
\nopagebreak

\def\oneh{{\textstyle {1\over 2}}}
\def\oneq{{\textstyle {1\over 4}}}
\def\onetre{{\textstyle {1\over 3}}}
\def\onesix{{\textstyle {1\over 6}}}
\def\onesixteen{{\textstyle {1\over 16}}}
\def\treh{{\textstyle {3\over 2}}}
\def\treq{{\textstyle {3\over 4}}}
\def\trecinq{{\textstyle {3\over 5}}}
\def\twothree{{\textstyle {2\over 3}}}

\def\eq{\begin{equation}}
\def\ee{\end{equation}}
\def\eqa{\begin{eqnarray}}
\def\eea{\end{eqnarray}}

\def\bra#1{\mbox{$\langle #1\vert $}}
\def\ket#1{\mbox{$\vert #1\rangle$}}

\begin{document}


\centerline{\Large{\bf Photocouplings of nucleon resonances}}

\medskip

\centerline{\Large{\bf with gluon and pion exchange currents}}

\vskip 1.5cm

\centerline{\large{E.~Perazzi$^{1}$, M.~Radici$^{1,2}$ and
S.~Boffi$^{1,2}$}}

\vskip 1.0cm

\centerline{\small $^1$~Dipartimento di Fisica Nucleare e Teorica,
Universit\`a di Pavia, and}

\centerline{\small $^2$~Istituto Nazionale di Fisica Nucleare, 
Sezione di Pavia, Pavia, Italy}

\vskip 1.5cm


\begin{abstract}

\noindent Photocouplings for the nucleon resonances most relevant to total
photoabsorption are calculated in a constituent quark model including gluon 
and pion exchange. The sensitivity of the results to the different 
ingredients of the model is discussed.

\end{abstract}

\bigskip



\vskip 1.5cm
\clearpage


\begin{section}{Introduction}

The structure and dynamics of hadrons have to be understood in
terms of a theoretical description of strong interactions ultimately
derived from the theory of quantum chromodynamics (QCD), but in a regime
where the underlying QCD dynamics cannot be treated perturbatively. Thus,
the complexity of hadrons requires the development of soluble models not
only retaining the basic symmetries of QCD but at the same time also
containing the relevant dynamical features of interacting quarks and
gluons. Quark models, i.e. models where the effective degrees of freedom
are massive quarks moving in a long-range confining potential and
interacting through two-body forces, have been very successful in
explaining spectroscopic properties of hadrons (see, e.g.,
refs.~\cite{[Close],[Bhaduri]}). In fact, the basic SU(6)
spin-flavour symmetry represents a stringent constraint on the
classification of baryon resonances in terms of multiplets. Some residual
interaction is then responsible for the ordering and mass splitting of
baryon states. 

In particular, according to the analysis of ref.~\cite{[DeRujula]} in the
nonrelativistic constituent quark model (CQM) (see, e.g.,
refs.~\cite{[Isgura],[Isgurb],[Karl]} and references therein) the residual
interaction is identified with the hyperfine-like part of the
nonrelativistic reduction of the one-gluon exchange diagram. It contains a
contact and a tensor term arising from the interaction energy of
two colour-magnetic dipoles. Other terms such as the spin-orbit part and
the momentum dependent terms are usually
neglected~\cite{[isgurkarluno],[isgurkarldue],[isgurkarltre],%
[isgurkarlquattro]}
on the ground that momentum dependent terms are of higher order in the
nonrelativistic reduction and the Thomas precession by the confining
forces tends to cancel colour-magnetic spin-orbit terms. Indeed, there is
spectroscopic evidence of small spin-orbit effects, while the hyperfine
interaction plays an essential role giving rise to a (negative) charge
radius for the neutron. 

On the other hand, the spectra of the confirmed states of the nucleon and
the $\Lambda$ hyperon split into a low energy sector of well separated
states without nearby parity partners, and a high energy sector with an
increasing number of near parity doublets. This fact has recently been
interpreted~\cite{[Riska],[Glozman],[Glozmanb]} as a piece of evidence
that the approximate chiral simmetry of QCD is realized in the hidden
Nambu-Goldstone mode at low excitation and in the explicit Wigner-Weyl mode
at high excitation. As a consequence, at distances beyond that of
spontaneous chiral symmetry breaking (0.2--0.3 fm) the appropriate
effective degrees of freedom for the description of the baryon structure
should be the constituent quarks with complicated internal structure and the
chiral meson fields. Besides the central confining part of the quark-quark
interaction one should then assume a chiral interaction that is mediated by the
octet of pseudoscalar mesons between the constituent quarks. The perturbative
gluon exchange interaction would become important only at length scales smaller
than that of the spontaneous chiral symmetry breaking. This model leads to
orderings of the positive and negative energy states in the baryon spectra
which agree with the observed ones in all
sectors~\cite{[Riska],[Glozman],[Glozmanb]}. The overall small spin-orbit
splitting in the baryon spectrum can also be qualitatively explained by the
absence of any spin-orbit component in the pseudoscalar exchange interaction.

The problem whether mesons should also be included among the effective
degrees of freedom in the baryon dynamics was already considered in the
literature~\cite{[Ichimura],[Rho],[Beyer],[DG],[WW],[BYIF],[YBFA],[Robson]}.
In particular, pion fields are introduced to restore chiral symmetry in
the nonrelativistic quark model. They demand that pion exchange currents
be included in order to account for partially conserved axial currents and
electromagnetic current conservation~\cite{[Robson]}.

However, to understand the internal dynamics and to test different aspects
of a model it is also necessary to gain information on the different
components of the baryon wave functions. This can fruitfully be achieved in
particular by looking at the electromagnetic form factors and the
transition (helicity) amplitudes in electromagnetic excitation of the
baryon resonances. A consistent analysis of the effects of gluon and pion
exchange currents on the nucleon electromagnetic form factors has been done
in refs.~\cite{[Buchmannfirst],[Buchmann]}. Helicity amplitudes have been
studied in a variety of
models~\cite{[Copley],[isgurkarlkoniuk],[Koniuk],[Stancu],[CloseLi],%
[LiClose],[Capstick],[CKeister],[Warns],[Warnsb],[Bijker]},
but the excitation or decay of a baryon has always been assumed to proceed
through a single quark transition even when considering three-body
forces~\cite{[Ferraris],[Aiello]}. Only in a few cases~\cite{[WW],[Robson]}
two-body currents arising from pion exchange have been considered.

A general feature of the results obtained for the helicity amplitudes is
the possibility of reproducing the correct sign and the relative strength
of the photocouplings for the majority of low-lying baryon resonances.
The main discrepancies, apart from the otherwise puzzling Roper resonance,
arise in the underestimated strength of the P$_{33}$(1232), D$_{13}$(1520)
and F$_{15}$(1680) states. These are the states mainly responsible for the
three well known peaks in the total photoabsorption cross
section~\cite{[Armstrong],[MacCormick]} corresponding to M1, E1 and E2
absorption, respectively~\cite{[brizzo],[brizzobis]}. Therefore their good
description is one of the first requests to any model.

In this paper the role of gluon and pion exchange currents is
investigated with reference to the photon decay of the P$_{33}$(1232),
D$_{13}$(1520) and F$_{15}$(1680) resonances within the constituent quark
model. In sect. 2 some features of the model are reviewed and the wave
functions describing these states are defined. In sect. 3 the expressions
for the two-body currents and details about the calculations are given. 
The results are discussed in sect. 4 and concluding remarks are presented
in the final section.

\end{section}

\begin{section}{Configuration mixing in the constituent quark model}

In the CQM a baryon is described as a nonrelativistic three-quark system in
which quarks interact via two-body potentials simulating the main features
of QCD. In the case of three equal masses the Hamiltonian is given by
\eq
H = \sum_{i=1}^3\left( m_{\rm q} + \frac{{\vec p}_i^2}{2m_{\rm q}}\right) 
- \frac{{\vec P}^2}{6m_{\rm q}} + \sum_{i<j}^3 V_{\rm conf}({\vec
r}_i,{\vec r}_j) + \sum_{i<j}^3 V_{\rm res}({\vec r}_i,{\vec r}_j) ,
\label{eq:hamiltonian}\\ \nonumber
\ee
where $m_{\rm q}=\onetre M_{\rm N}=313$ MeV is the constituent quark mass,
${\vec P}$ is the total momentum of the baryon and ${\vec r}_i$, ${\vec
p}_i$ are the spatial coordinate and momentum of the $i$th quark,
respectively. The first three terms describe the intrinsic nonrelativistic
free motion of the quarks including their mass. The fourth term is a two-body
confinement potential, which is usually taken to have the harmonic oscillator
form: \eq
V_{\rm conf}({\vec r}_i,{\vec r}_j) = - a_{\rm
c}{\vec\lambda}_i\cdot{\vec\lambda}_j ({\vec r}_i - {\vec r}_j)^2 ,
\label{eq:confinement}\\ \nonumber
\ee
where ${\vec\lambda}_i$ is the SU(3) colour matrix. With the expectation
value $\langle\oneh{\vec\lambda}_i\cdot\oneh{\vec\lambda}_j\rangle =
-\twothree$ in a baryon, the relation between the oscillator
frequency $\omega$ and $a_{\rm c}$ is given by $a_{\rm c}=\onesixteen
m_{\rm q}\omega^2$. The oscillator spring constant is defined as
$b=\sqrt{\hbar/m_{\rm q}\omega}\equiv \alpha^{-1}$. All together these
terms (apart from the total rest mass) define the unperturbed Hamiltonian
$H_0$, while $V_{\rm res}$ describes the residual interaction.

Introducing the new (Jacobi) variables
\eq
{\vec\rho} = \frac{1}{\sqrt{2}}({\vec r}_1 -{\vec r}_2),\qquad
{\vec\lambda} = \frac{1}{\sqrt{6}}({\vec r}_1 + {\vec r}_2 -
2 {\vec r}_3),
\label{eq:coordinates}\\ \nonumber
\ee
and their conjugate momenta
\eq
{\vec p}_\rho = \frac{1}{\sqrt{2}}({\vec p}_1 -{\vec p}_2),\qquad
{\vec p}_\lambda = \frac{1}{\sqrt{6}}({\vec p}_1 + {\vec p}_2 -
2 {\vec p}_3),
\label{eq:momenta}\\ \nonumber
\ee
the unperturbed Hamiltonian $H_0$ separates into two independent
oscillators with frequency $\omega$:
\eq
H_0 = \frac{{\vec p}_\rho^2}{2m_{\rm q}} + 
\oneh m_{\rm q}\omega^2{\vec\rho}^2 
+ \frac{{\vec p}_\lambda^2}{2m_{\rm q}} +  
\oneh m_{\rm q}\omega^2{\vec\lambda}^2 .
\label{eq:unperturbed}\\ \nonumber
\ee
The corresponding eigenstates are classified into multiplets according
to the underlying SU(6) spin-flavour symmetry. They are labelled as
$\ket{B\, ^{2S+1}X_J}_\pi$, where $B=N,\Delta$ denotes the SU(3)
representation (flavour octet or decuplet), $^{2S+1}X_J$ is the usual
spectroscopic notation [with $X=S,P,D,\dots$ staying for $L$ and $J$
($S$) for the total angular momentum (spin) of the state] and $\pi=A,M,S$
indicates the symmetry type of the SU(6) states (antisymmetric, mixed and
symmetric, respectively). 

Configuration mixing is introduced by the residual interaction. As a
consequence, in the $2\hbar\omega$ configuration space the nucleon and the
baryon resonances under study have the following expansion:  
\eqa 
\ket{N(940)} & = & a_S\ket{N\, ^2S_{1/2}}_S +
a'_S\ket{N\, ^2S'_{1/2}}_S  + a_M\ket {N\,^2S_{1/2}}_M \nonumber \\ 
& &{} + a_D\ket{N\, ^4D_{1/2}}_M + a_P\ket{N\, ^2P_{1/2}}_A, \nonumber \\
& & \nonumber \\ 
\ket{\Delta (1232)} & = &
b_S\ket{\Delta\, ^4S_{3/2}}_S + b'_S\ket{\Delta\, ^4S'_{3/2}}_S 
+ b_D\ket {\Delta\,^4D_{3/2}}_S + b'_D\ket{\Delta\, ^2D_{3/2}}_M,
\nonumber \\
& & \nonumber \\
\ket{N(1520)} & = &
c_P\ket{N\, ^2P_{3/2}}_M + c'_P\ket{N\, ^4P_{3/2}}_M ,
\nonumber \\
& & \nonumber \\
\ket{N(1680)} & = &
c_D\ket{N\, ^2D_{5/2}}_S + c'_D\ket{N\, ^2D_{5/2}}_M +
c''_D\ket{N\, ^4D_{5/2}}_M .
\label{eq:wavefunct}\\ \nonumber
\eea
According to ref.~\cite{[DeRujula]} the simplest quark interaction is due
to the nonrelativistic reduction of the one-gluon exchange, but usually
only the spin-spin hyperfine interaction is retained. The expansion
coefficients of the above configuration mixing are then obtained by
diagonalizing the hyperfine interaction. 

On the other hand, also pion degrees of freedom should play an important
role. The one-pion exchange potential (OPEP) can be considered as the
first term in the flavour structure of the chiral interaction that
is mediated by the octet of pseudoscalar mesons, as proposed, e.g.,
in refs.~\cite{[Riska],[Glozman],[Glozmanb]} to account for the correct
ordering of positive and negative parity states in the baryon spectrum. In
fact, the one-pion exchange is the dominant residual interaction in the
calculation of the $\Delta$--N mass splitting and the neutron charge
radius~\cite{[Buchmannfirst],[Buchmann]}. 

Different admixture coefficients are obtained by diagonalizing the total
Hamiltonian containing the one-pion  exchange potential as well as the whole
one-gluon exchange potential. When only the hyperfine interaction without
spin-orbit contribution is considered, the fifth state $\ket{N\, ^2P_{1/2}}_A$
in the nucleon case is left uncoupled. In general, this is not the case for the
total Hamiltonian here considered~\cite{[Buchmannfirst],[Buchmann]}, but in a
test calculation its effect was found rather small. Therefore, it will be
disregarded in the following.

Many solutions to eq. (\ref{eq:wavefunct}) are available in the literature
depending on the choice of the involved parameters. Some of them are reported
in Table 1 for reference in the discussion of our results. Column KI refers to
the mixing angles derived in the original Isgur--Karl model as quoted in
ref.~\cite{[Warns]} and used, e.g., for the excited states in the calculation
of baryon decays in ref.~\cite{[Koniuk]}. Column G is an almost equivalent
configuration mixing obtained within the same model~\cite{[Mauro]} but with a
different choice of the parameters. Column WW is the corrected result of
ref.~\cite{[WW]} (as reported in ref.~\cite{[Buchmannfirst]}). A static pion
exchange potential (including the zero-range OPEP) and a
simplified colour hyperfine interaction were adopted in the diagonalization of
the whole Hamiltonian. Columns B1 and B2 are the results of
ref.~\cite{[Buchmann]} with and without OPEP,
respectively. The whole Hamiltonian of ref.~\cite{[Buchmann]} includes a
two-body confinement potential, the complete one-gluon exchange potential
with spin-orbit terms and, when present, the one-pion exchange potential.
In the latter the finite size of the pion-quark vertex was also taken into
account. In the original model of ref.~\cite{[Buchmannfirst],[Buchmann]}
only the nucleon and the P$_{33}(1232)$ mixing coefficients were
calculated. The coefficients in Table 1 for the D$_{13}(1520)$ and
F$_{15}(1680)$ resonances are an arbitrary guess inspired by the fact that
diagonalizing the whole Hamiltonian reduces the size of the larger
coefficient of the baryon wave function in the expansion on the harmonic
oscillator basis with respect to the case where the residual interaction
is treated perturbatively. The rational for using this guess is to test the
sensitivity of the results to configuration mixing.

\end{section}

\begin{section}{The electromagnetic currents}

As usual, the nonrelativistic one-body current is the sum of the
convective and spin-magnetic terms, i.e.
\begin{equation}
{\vec J}_{\rm imp}({\vec r}_i,{\vec q}) = \frac{e_i}{2m_{\rm q}}
\{{\vec p}_i,\exp({\rm i}{\vec q}\cdot {\vec r}_i)\} +
\mu_i \, {\rm i} \, [{\vec\sigma}_i\times {\vec p}_i,\exp({\rm i}{\vec
q}\cdot{\vec r}_i)] , 
\label{eq:jimp}\\ \nonumber
\end{equation}
where ${\vec q}$ is the external photon three-momentum and the
$i$th quark has charge $e_i = \onesix (1 + 3 \tau_{iz}) e$ 
and magnetic moment $\mu_i = g e_i/2m_{\rm q}$. An anomalous quark magnetic
moment implies $g\ne 1$, but here $g=1$ has been assumed.

The two-body current depends on the assumed baryon Hamiltonian. It is
basically determined by gluon and pion exchanges between quarks. In the
approach of refs.~\cite{[Buchmannfirst],[Buchmann]} they are given by the
following contributions
\begin{equation}
{\vec J}_{\rm gq{\overline q}}({\vec r}_i,{\vec r}_j,{\vec
q}) = 
- \frac{\alpha_{\rm s}}{4m^2_{\rm q}}{\vec\lambda}_i\cdot{\vec\lambda}_j
\frac{1}{r^3}
\left[e_i\,  \exp({\rm i}{\vec q}\cdot{\vec r}_i)\,  \oneh({\vec\sigma}_i
+ {\vec\sigma}_j)\times{\vec r} + (i\leftrightarrow j)\right],
\label{eq:jgqq}\\ \nonumber
\end{equation}
\begin{eqnarray}
{\vec J}^{\,\rm IS}_{\pi\rm q{\overline q}}({\vec r}_i,{\vec r}_j,{\vec
q}) & = & \frac{{\rm i}e}{6} \frac{f_{\pi{\rm q}}^2}{4\pi m_\pi^2}
\frac{\Lambda^2}{\Lambda^2 - m^2_\pi}\frac{1}{2m^2_{\rm q}}
{\vec\tau}_i\cdot{\vec\tau}_j \nonumber \\
& & \nonumber \\
& & \times \left[\exp({\rm i}{\vec q}\cdot{\vec r}_i)\,
\left({\vec q}\times{\vec\nabla}_r \right)\,
\left({\vec\sigma}_j\cdot{\vec\nabla}_r\right)
+ (i\leftrightarrow j)\right] \nonumber \\
& & \nonumber \\
& & \times \left[\frac{\exp(-m_\pi r)}{r} - 
\frac{\exp(-\Lambda r)}{r}\right], \label{eq:jpiqqis}\\ \nonumber
\end{eqnarray}
\begin{eqnarray}
{\vec J}^{\,\rm IV}_{\pi\rm q{\overline q}}({\vec r}_i,{\vec r}_j,{\vec
q}) & = & e \frac{f_{\pi{\rm q}}^2}{4\pi m_\pi^2}
\frac{\Lambda^2}{\Lambda^2 - m^2_\pi} \nonumber \\
& & \nonumber \\
& & \times \left[ [{\vec\tau}_i\times{\vec\tau}_j]_z \,
\exp({\rm i}{\vec q}\cdot{\vec r}_i)\,
{\vec\sigma}_i\,\left({\vec\sigma}_j\cdot{\vec\nabla}_r \right) 
+ (i\leftrightarrow j)\right] \nonumber \\
& & \nonumber \\
& & \times \left[\frac{\exp(-m_\pi r)}{r} - 
\frac{\exp(-\Lambda r)}{r}\right], \label{eq:jpiqqiv}\\ \nonumber 
\end{eqnarray}
\begin{eqnarray}
{\vec J}_{\gamma\pi\pi}({\vec r}_i,{\vec
r}_j,{\vec q}) & = & e \frac{f_{\pi{\rm q}}^2}{4\pi m_\pi^2}
\frac{\Lambda^2}{\Lambda^2 - m^2_\pi}
[{\vec\tau}_i\times{\vec\tau}_j]_z\, 
\left({\vec\sigma}_i\cdot{\vec\nabla}_i\right)\,
\left({\vec\sigma}_j\cdot{\vec\nabla}_j\right)\nonumber \\
& & \nonumber \\
& & \times\int_{-1/2}^{1/2}{\rm d}v\,\exp\left[ {\rm i}{\vec q}\cdot({\vec R}
-{\vec r}v) \right] \nonumber \\
& & \nonumber \\
& & \times
\left[{\vec z}_\pi \frac{\exp(-L_\pi r)}{L_\pi r} - 
{\vec z}_\Lambda\frac{\exp(-L_\Lambda r)}{L_\Lambda r}\right],
\label{eq:jgammapipi}\\ \nonumber 
\end{eqnarray}
where $m_\pi$ is the pion mass, ${\vec r} = {\vec r}_i - {\vec r}_j$,
${\vec R} = \oneh({\vec r}_i+{\vec r}_j)$, ${\vec z}_m = L_m{\vec r} +{\rm
i}vr{\vec q}$ and $L_m(q,v) = [\oneq q^2(1-4v^2) + m^2]^{1/2}$. The
pion-quark cutoff mass $\Lambda$ describes the size of the pion-quark
interaction region. For $\Lambda\to\infty$ the full $\delta$-function
interaction is recovered in the one-pion exchange potential.

Eqs.~(\ref{eq:jgqq})--(\ref{eq:jpiqqiv}) describe quark-antiquark
pair creation processes induced by the external photon with subsequent
annihilation of the quark-antiquark pair into a gluon (gluon pair
current) or a pion (pion pair current) which is then absorbed at the
site of another quark. The pion pair current is divided into its
isoscalar (IS) and isovector (IV) components. Eq.~(\ref{eq:jgammapipi})
describes the pionic current where the photon couples to the pion
directly. Gauge invariance of the current operators is preserved up to 
the considered order $1/m^2_{\rm q}$ in the nonrelativistic 
reduction~\cite{[Buchmann]}.

The quark-gluon fine-structure constant $\alpha_s$ was adjusted in
refs.~\cite{[Buchmannfirst],[Buchmann]} to obtain the correct $\Delta$--N
mass splitting through the relation
\begin{equation}
\alpha_s = \treq\, \sqrt{2\pi} \, m_{\rm q}^2 \, b^3 \, 
[M_\Delta - M_{\rm N} - \delta_\pi(b)] ,
\label{eq:alfas}\\ \nonumber 
\end{equation}
where $\delta_\pi(b)$ is the pion contribution to the $\Delta$--N mass
splitting. Thus, $\alpha_s$ and the oscillator spring constant
$b=\alpha^{-1}$ depend on the presence of pion exchanges and the
possibility of configuration mixing. The pion-quark coupling 
constant $f^2_{\pi{\rm q}}/4\pi$ is related to the pion-nucleon 
coupling constant $f^2_{\pi{\rm N}}/4\pi = 0.08$ via 
$f_{\pi{\rm q}} = \trecinq f_{\pi{\rm N}}$~\cite{[Ito]}. 
The model parameters derived in ref.~\cite{[Buchmann]} are summarized in
Table 2 together with another option related to ref.~\cite{[WW]}.

Since the isospin operator ${\vec\tau}_i\cdot{\vec\tau}_j$ cannot 
connect states with different isospin, the contribution of the isoscalar 
pion pair current to the excitation of the ${\rm P}_{33}$ resonance 
vanishes identically. The operator 
${\vec J}^{\,\rm IS}_{\pi {\rm q {\overline q}}}$ contains an 
additional $1/m^2_{\rm q}$ term with respect to 
${\vec J}^{\, \rm IV}_{\pi {\rm q {\overline q}}}$ 
(see Eqs.~(\ref{eq:jpiqqis}) and (\ref{eq:jpiqqiv})) so that its contribution 
to the excitation of the other two resonances is anyway smaller than that of 
${\vec J}^{\, \rm IV}_{\pi {\rm q {\overline q}}}$. In a test calculation 
this contribution also turned 
out to be much smaller than those coming from the 
other pieces of the current operator. Therefore, in the following 
${\vec J}^{\,\rm IS}_{\pi {\rm q {\overline q}}}$ will be 
neglected. The same test was performed with no configuration mixing 
on the pionic current ${\vec J}_{\gamma\pi\pi}$, which 
is more involved because of the additional integration required. 
As already found in ref.~\cite{[Robson]}, this piece also gives a
negligible contribution and will be disregarded. Consequently, the 
interaction Hamiltonian $H_{\rm int}$ related to two-body 
currents becomes 
\begin{eqnarray}
H_{\rm int} &\simeq &- \sum_{i<j}^3 \left[ 
{\vec J}_{\rm gq{\overline q}}({\vec r}_i,{\vec r}_j,{\vec q}) 
\cdot {\vec A} \, + \, 
{\vec J}^{\,\rm IV}_{\pi\rm q{\overline q}}({\vec r}_i,{\vec r}_j,
{\vec q}) \cdot {\vec A} \right] \nonumber \\
& & \nonumber \\
&= &- 3 \  \left[ 
{\vec J}_{\rm gq{\overline q}}({\vec r}_1,{\vec r}_2,{\vec q}) 
\cdot {\vec A} \, + \, 
{\vec J}^{\,\rm IV}_{\pi\rm q{\overline q}}({\vec r}_1,{\vec r}_2,
{\vec q}) \cdot {\vec A} \right] \nonumber \\ 
& & \nonumber \\
&\equiv &\, H_{{\rm gq \overline{\rm q}}} \, + \, 
H_{\pi{\rm q \overline{\rm q}}}, 
\label{eq:hamint}\\ \nonumber
\end{eqnarray}
where $\vec A$ is the external photon field and symmetry 
properties of the operators in 
eqs.~(\ref{eq:jgqq}) and (\ref{eq:jpiqqiv}) have been used. By 
considering the symmetries of the various matrix elements of 
$H_{\rm int}$, i.e. by also taking into account the symmetries 
of the baryon wave function, one finds that the exchange 
term in eqs.~(\ref{eq:jgqq}) and (\ref{eq:jpiqqiv}) equals the 
direct one. Therefore, the two contributions 
$H_{{\rm gq \overline{\rm q}}}, H_{\pi{\rm q \overline{\rm q}}}$ 
actually involve just 
the direct term in eqs.~(\ref{eq:jgqq}) and (\ref{eq:jpiqqiv}) 
multiplied by 6
and can be written in terms of the Jacobi coordinates of 
eq. (\ref{eq:coordinates}) as
\begin{eqnarray}
H_{{\rm gq \overline{\rm q}}} &= &\frac{{\rm i}\alpha_{\rm s}}
{m^2_{\rm q}}\, \frac{\sqrt{2}}{\rho^3} \  e_1 \,  
\exp\left[ {\rm i}q\left(\frac{\rho_z}{\sqrt{2}} + \frac{\lambda_z}
{\sqrt{6}}\right)\right] \nonumber \\
& & \nonumber \\
& &\times \left[ \frac{1}{\sqrt{2}} \left( 
\sigma_{1z} + \sigma_{2z}\right) \rho_+ - \left( \sigma_{1+} + 
\sigma_{2+} \right) \rho_z \right] ,\\
& & \nonumber \\
H_{\pi{\rm q \overline{\rm q}}} &= &\frac{6}{\sqrt{2}} \,   
\frac{ef^2_{\pi{\rm q}}}{4\pi m^2_{\pi}} \frac{\Lambda^2}
{\Lambda^2 - m^2_{\pi}} \ [\vec\tau_1 \times \vec\tau_2]_z \, 
\exp\left[{\rm i}q \left( \frac{\rho_z}{\sqrt{2}} + 
\frac{\lambda_z}{\sqrt{6}} \right)\right] \nonumber \\
& & \nonumber \\
& &\times \left( \sqrt{2} \, \sigma_{2+} \frac{\partial}{\partial 
\rho_+} + \sqrt{2} \, \sigma_{2-} \frac{\partial}{\partial \rho_-} 
+ \sigma_{2z} \frac{\partial}{\partial \rho_z} \right) \  
\sigma_{1+} \nonumber \\
& & \nonumber \\
& &\times \left[ \frac{\exp(- \sqrt{2} m_{\pi} \rho)}{\rho} - 
\frac{\exp(- \sqrt{2} \Lambda \rho)}{\rho} \right]. 
\label{eq:hgqqpiqq} \\ \nonumber
\end{eqnarray}

By inspecting the isospin part of the SU(6) baryon eigenstates, 
it is possible to define some general selection rules on the transition 
matrix elements between the nucleon and the three considered resonances, 
${\rm P}_{33}(1232)$, ${\rm D}_{13}(1520)$ and ${\rm F}_{15}(1680)$. In fact,
the isospin part  both of the current operator and the baryon wave function is
factorized and  the spatial and spin structures of the proton and neutron wave
functions  are equal. Therefore, only the isospin transition matrix elements
can produce  different results. The SU(6) eigenstates with isospin $I=\oneh$
can be cast in  two different kinds of states with mixed symmetry as 
\begin{equation}
\vcenter{\openup1\jot \tabskip=0pt plus1fil  \halign
{\tabskip=0pt
  $\hfil#$ &$#$&$#$
 &$\hfil{}#{}$ &$#$&$#$\tabskip=0pt plus1fil&
  \llap{#}\tabskip=0pt\cr
 I_z &=& \oneh  & I_z &=& -\oneh   \cr
\ket{M_{\rho}}&=&{\textstyle{1\over\sqrt{2}}}\left( {\rm udu-duu}\right), \quad&
\ket{M_{\rho}}&=&{\textstyle{1\over\sqrt{2}}}\left( {\rm udd-dud}\right),\cr
\ket{M_{\lambda}}&=&-{\textstyle{1\over\sqrt{6}}}\left( {\rm udu+duu-2uud}\right), \quad&
\ket{M_{\lambda}}&=&{\textstyle{1\over\sqrt{6}}}\left( {\rm udd+dud-2ddu}\right), \cr
}} \label{eq:su6iso1}
\end{equation}
while the symmetric eigenstates with $I=\treh$ are
\begin{equation}
\vcenter{\openup1\jot \tabskip=0pt plus1fil  \halign
{\tabskip=0pt
  $\hfil#$ &$#$&$#$
 &$\hfil{}#{}$ &$#$&$#$\tabskip=0pt plus1fil&
  \llap{#}\tabskip=0pt\cr
 I_z &=& \oneh  & I_z &=& -\oneh  \cr
\ket{S}&=&{\textstyle{1\over\sqrt{3}}}\left( {\rm udu+duu+uud}\right), \qquad &
\ket{S}&=&{\textstyle{1\over\sqrt{3}}}\left( {\rm udd+dud+ddu}\right).\cr
}} \label{eq:su6iso3} 
\end{equation}
Here, u (d) denotes the isospin up (down) quark state. 

In the case of the transition to the ${\rm P}_{33}(1232)$ resonance one deals
with the isospin matrix elements  
$\bra{S} \hat T \ket{M_{\rho}}$, $\bra{S} \hat T \ket{M_{\lambda}}$,  where
$\hat T = \tau_{1z}$,  $[\vec \tau_1 \times  \vec \tau_2]_z$. From
eqs.~(\ref{eq:su6iso1}) and (\ref{eq:su6iso3}) it is easy  to verify that
for every piece of the current operator the matrix elements  involving an
initial neutron equal the corresponding ones with an initial  proton. In the
case of the transition to the ${\rm D}_{13}(1520)$ and ${\rm F}_{15}(1680)$ 
resonances similar arguments lead to the conclusion that the pion-exchange 
transition amplitudes with an initial neutron are opposite to the
corresponding  ones with an initial proton, because the involved isospin matrix
elements are  $\bra{M_{\rho}} [\vec \tau_1 \times \vec \tau_2]_z
\ket{M_{\lambda}}$. The  one-body and gluon-exchange operators will generally
give different results.

\end{section}

\begin{section}{Results}

Calculations have been performed for the helicity amplitudes for photon
decay of the three resonances which are most excited in total
photoabsorption, i.e. P$_{33}(1232)$, D$_{13}(1520)$ and F$_{15}(1680)$.
Analytical formulae are obtained for the contribution of the one-body current.
A numerical integration over $\rho$ is further required for the matrix
elements of the two-body currents after some effort in reducing the analytical
expression.

Let us first consider the one-body current only and the corresponding
results for the calculated photocouplings shown in Table 3. As usual,
here and in the following a factor $+{\rm i}$ in front of all D$_{13}(1520)$
amplitudes has been suppressed. With the simplest assumption, i.e. the three
quarks in the baryon in the lowest $(0s)^3$ harmonic oscillator state, the
results are strongly dependent on the oscillator spring constant. In agreement
with previous suggestions~\cite{[Copley],[Koniuk]} we choose $\alpha = 0.41$
GeV $=2.0778$ fm$^{-1}$. Thus, the helicity amplitude $A^p_{1/2}$ for photon
decay of the F$_{15}(1680)$ resonance to the proton is almost vanishing.
The effects of configuration mixing are sizeable, particularly when also the
nucleon ground state is mixed as can be realized by comparing columns KI and
G1. The differences between columns G1 and G2 depend on the oscillator spring
constant~\cite{[Mauro],[brizzo]}. The first choice (G1) is in agreement with
the suggestion of refs.~\cite{[Copley],[Koniuk]} and gives mixing coefficients
G similar to KI (see Table 1). The same mixing is used for G2, but the
oscillator spring constant is fixed by the proton r.m.s. radius. The well known
difficulty of simultaneously reproducing static and dynamic properties of the
baryons is here confirmed and can be ascribed to a confinement radius ($\sim
0.5$ fm) much less than the nucleon r.m.s. radius. In any case, the best result
(G1) clearly still underestimates the size of the photocouplings. The general
trend is to make the comparison with experimental data worse than the results
without configuration mixing.

In Table 4 results from a variety of differently related models are collected
for comparison. SS~\cite{[Stancu]} refers to the same nonrelativistic one-body
current operator but with a more sophisticated (correlated) basis for the wave
functions derived variationally in the frame of a flux-tube quark
model~\cite{[Panda]}. Relativistic corrections to the transition operator
have been introduced in refs.~\cite{[CloseLi],[LiClose]} and
refs.~\cite{[Warns],[Warnsb]} while keeping the Isgur--Karl model wave
functions and in Table 4 the corresponding results are labelled LC and WPR,
respectively. The results of the relativized quark model of
ref.~\cite{[Capstick]} are indicated by C. The photocouplings obtained in
the algebraic approach of ref.~\cite{[Bijker]} and with a hypercentral
potential in a three-body force model~\cite{[Ferraris],[Aiello]} are also
given as BIL and Hyp, respectively. Since the helicity amplitudes calculated
with the one-body current are dominated by the spin-flavour structure of the
states~\cite{[Aiello]}, it is not surprising that all the above models give
qualitatively the same results.

Just to have an idea about the consequences of the results shown in Table 3
on some observables, one can calculate the total and backward cross
sections for $\pi^0$ photoproduction. According to ref.~\cite{[Copley]} they
are given in terms of the helicity amplitudes $A_J$ by
\eq
\sigma_T = K_I\,
\frac{M_{\rm N}}{M_{\rm R}}\,\frac{x}{\Gamma} \,
2\left[ \vert A_{1/2}\vert^2 + \vert A_{3/2}\vert^2\right] ,
\label{eq:totale}\\ \nonumber 
\ee
\eq
\left.\frac{{\rm d}\sigma}{{\rm d}\Omega}\right\vert_{\theta=\pi} =
K_I\,
\frac{2J+1}{4\pi}\,\frac{M_{\rm N}}{M_{\rm R}}\,\frac{x}{\Gamma}
\,\vert A_{1/2}\vert^2, 
\label{eq:backward}\\ \nonumber 
\ee
respectively. In eqs. (\ref{eq:totale}) and (\ref{eq:backward}) $x$ and
$\Gamma$ are the elasticity and total width of the resonance of mass 
$M_{\rm R}$ and the factor $K_I = \onetre,\twothree$ corresponds to isospin
$I=\oneh,\treh$, respectively. Apparently, according to the results
shown in Table 5 configuration mixing does not help to improve the
agreement with data and in any case the results are quite sensitive to
the choice of the oscillator spring constant. 

Let us now consider the effect of two-body currents. In Table 6 results
are given for the photon decay of the P$_{33}(1232)$ resonance. The values
labelled B1 and B2 are obtained using the configuration mixing and the model
parameters of ref.~\cite{[Buchmann]} with and without taking into account
pion exchanges, respectively. They are compared with results labelled WW and
calculated with the configuration mixing and the model parameters taken
from the simplified model of ref.~\cite{[WW]}. For reference, the photocouplings
calculated in ref.~\cite{[Robson]} considering pion exchange only and no
configuration mixing are also shown. Quite similar results are obtained
when performing a complete calculation irrespective of the model parameters
(B1 and WW) or without configuration mixing but including OPEP (R). A dramatic
effect comes in when switching pion exchange off as in B2. In order to
illustrate this effect, the contributions to the helicity amplitudes coming
from the different pieces of the electromagnetic current are separately shown
in Table 7 with and without configuration mixing. Of course, the main
contribution is always given by the one-body current. Its size is almost the
same in the different cases showing that the results are stable when
consistently using wave functions and model parameters. Gluon and pion exchange
currents always enter with the opposite sign, as also noticed in
refs.~\cite{[Buchmannfirst],[Buchmann]} when calculating the charge and
magnetic form factors of the nucleon. However, since the pionic current is
quite negligible in the calculation of the photocouplings, the partial
cancellation between the isovector pion pair and pionic currents found in
ref.~\cite{[Buchmann]} for the magnetic moments does not occur here. As a
consequence, the pion exchange contribution is rather large and its sign is
responsible for the dramatic reduction effect on the size of the photocouplings.
This leads to results in contrast with the quoted experimental values. The same
effect was found in the simplified approach to the pion-exchange current
contribution of ref.~\cite{[Robson]} with no configuration mixing.
Configuration mixing reduces the size of gluon exchange and increases that of
pion exchange, thus enlarging the discrepancy of the total result with
respect to the experimental value. When pion exchange is neglected in the
original Hamiltonian and the helicity amplitudes are consistently calculated
within the model (B2 in Tables 6 and 7) values much closer to the data are
obtained. 

A similar analysis can be extended to the other resonances. However, in
this case the mixing coefficients entering the wave functions have not yet
been derived consistently within the model of
refs.~\cite{[Buchmannfirst],[Buchmann]}. As often done in the past, due to
the lack of a better approximation we use the wave functions
labelled G in Table 1 and calculated in the Isgur--Karl model. The results are
shown in Table 8 with and without pion exchange currents (B1--G and B2--G,
respectively). Also shown in Table 8 are results derived without configuration
mixing (B1--(0s)$^3$ and B2--(0s)$^3$, respectively) and with two other
options of configuration mixing (B1--B1 and B2--B2) which are here proposed in
Table 1 with an asterisk in columns B1 and B2, respectively. With the exception
of the $A_{3/2}$ amplitudes of the F$_{15}(1680)$ resonance, similar results
are obtained without configuration mixing and with the configuration mixing G. 
A much larger effect comes from reducing the weight of the dominant state in the
expansion of the wave function, as proposed by options B1 and B2. Apart from
the $A^p_{1/2}$ amplitude and in contrast with the outlined general trend, a
better agreement with data is reached for the D$_{13}(1520)$ helicity amplitudes
also including OPEP, but the quantitative description of the photocouplings
within the model remains an open problem.

As a consequence of the above results, the values of the total and 
backward cross section for $\pi^0$ photoproduction of Table 3 change to
those reported in Table 9.

\end{section}

\begin{section}{Concluding remarks}

The effects of including two-body currents in the electromagnetic transition
operator have been investigated and photocouplings of some important nucleon
resonances calculated. The model Hamiltonian for the baryons is based
on the constituent quark model and includes gluon and pion exchange between
quarks with parameters fixed in ref.~\cite{[Buchmann]} by fitting the mass
spectrum. Different options of such parameters are available depending on
whether configuration mixing and/or pion exchange are included in the
model. The corresponding electromagnetic current operator satisfies gauge
invariance within the adopted nonrelativistic approach. 

Results for the photocouplings obtained with only the one-body current are more
sensitive to the value of the oscillator spring constant $b=\alpha^{-1}$ than
to configuration mixing. As previously observed in the literature, a value of
$b$ in agreement with the confinement radius gives much better results for the
photocouplings than a value determined by the nucleon r.m.s. radius. However,
the agreement with data is still only qualitative. 

When including two-body currents the contribution of gluon exchange has the
opposite sign to that of pion exchange. However, the size of the pion-exchange
contribution is much larger and pushes in the wrong direction with respect
to data. This is at variance of what happens for the magnetic moments of the
nucleon where the pion cloud gives only a small correction~\cite{[Buchmann]}.
Configuration mixing even increases such discrepancy and better
results are generally obtained when pion exchange is omitted in the original
Hamiltonian. For the D$_{13}(1520)$ a warning is necessary because the recent
results obtained by investigating two-pion photoproduction~\cite{[Braghieri]}
shed some light on different mechanisms responsible for the second peak in the
total photoabsorption cross section~\cite{[Leonid],[Bianchi]}. When better
understood, these mechanisms could well modify the experimental values of the
D$_{13}(1520)$ photocouplings.

In conclusion, the present analysis and the persistent discrepancy
with data illustrate the limitations of nonrelativistic models based on
constituent quarks. 

\end{section}

\bigskip

One of us (S.~B.) acknowledges useful conversations with A.~Buchmann,
M.~M.~Giannini and D.~O.~Riska. 
 
\clearpage


\clearpage


\noindent Table 1. Mixing coefficients for some baryonic states. KI from
ref.~\cite{[Koniuk]}, G from ref.~\cite{[Mauro]}, WW from ref.~\cite{[WW]} (as
reported  in ref.~\cite{[Buchmannfirst]}), B1 and B2 from
ref.~\cite{[Buchmann]}  with and without pion exchange, respectively.
Coefficients marked by an asterisk are a proposed guess.

\medskip

\begin{tabular}{crrrrr} \hline
\multicolumn{1}{c}{}
&\multicolumn{1}{c}{KI} 
&\multicolumn{1}{c}{G} 
&\multicolumn{1}{c}{WW} 
&\multicolumn{1}{c}{B1} 
&\multicolumn{1}{c}{B2} \\
\hline
N$(1232)$ &&&&& \\
$a_S$   &$0.930$  &$0.931$  &$0.906$  &$0.8210$  &$0.8301$   \\
$a'_S$  &$-0.290$ &$-0.274$ &$-0.383$ &$-0.5591$ &$-0.5460$  \\
$a_M$   &$-0.230$ &$-0.233$ &$-0.178$ &$-0.1051$ &$-0.1029$  \\
$a_D$   &$-0.040$ &$-0.067$ &$-0.045$ &$-0.0345$ &$-0.0258$  \\
\hline
P$_{33}(1232)$ &&&&& \\
$b_S$   &$0.970$  &$0.963$  &$0.994$  &$0.8798$  &$0.8900$   \\
$b'_S$  &$0.200$  &$0.231$  &$0.024$  &$-0.4692$ &$-0.4500$  \\
$b_D$   &$-0.100$ &$-0.119$ &$-0.090$ &$-0.0591$ &$-0.0565$  \\
$b'_D$  &$0.070$  &$0.075$  &$0.056$  &$0.0482$  &$0.0466$   \\
\hline
D$_{13}(1232)$ &&&&& \\
$c_P$   &$0.994$  &$0.994$  &$$        &$0.9000^*$  &$0.9000^*$     \\
$c'_P$  &$0.110$  &$0.111$  &$$        &$0.4359^*$  &$0.4359^*$  \\
\hline
F$_{15}(1232)$ &&&&& \\
$c_D$   &$0.880$  &$0.896$  &$$        &$0.8500^*$  &$0.8500^*$    \\
$c'_D$  &$-0.480$ &$-0.443$ &$$        &$-0.5000^*$ &$-0.5000^*$    \\
$c''_D$ &$0.010$  &$0.029$  &$$        &$0.1658^*$  &$0.1658^*$  \\
\hline
\end{tabular}

\vskip 2truecm

\noindent Table 2. Quark model parameters with regularized one-pion 
exchange ($\Lambda = 4.2$ fm$^{-1}$) and without pions (no $\pi$).
The parameters in columns $\pi$~(0s)$^3$ and no~$\pi$~(0s)$^3$ refer to
no configuration mixing, in columns B1 and B2 to configuration 
mixing~\cite{[Buchmann]}. In column WW the parameters of ref.~\cite{[WW]}
are given together with the regularized one-pion exchange.

\medskip

\leftskip 1truecm

\begin{tabular}{lrrrrr} \hline
\multicolumn{1}{c}{}
&\multicolumn{1}{c}{$\pi$ (0s)$^3$} 
&\multicolumn{1}{c}{no $\pi$ (0s)$^3$} 
&\multicolumn{1}{c}{B1} 
&\multicolumn{1}{c}{B2}
&\multicolumn{1}{c}{WW} \\
\hline
$\Lambda$ [fm$^{-1}$] &4.2 &  &4.2  &   &4.2\\
$b$ [fm]     &0.613 &0.603 &0.606 &0.576  &0.660\\
$\alpha^2$ [fm$^{-2}$] &2.661 &2.750 &2.723   &3.014 &2.295   \\
$\alpha_s$ &1.057 &1.543 &0.315 &0.720 &0.660 \\
\hline
\end{tabular}

\leftskip 0pt

\clearpage

\noindent Table 3. The helicity amplitudes $A_J$ calculated only with the
one-body current for photon decay into proton  (p) or neutron (n) in units
$10^{-3}$ GeV$^{-1/2}$. (0s)$^3$ without configuration  mixing; KI with
the configuration mixing quoted in Table 1, but only for the excited states; 
G1 and G2 with the complete configuration mixing G as quoted in Table 1;
$\alpha^2= 4.32$ fm$^{-2}$ in all cases, except G2 where $\alpha^2= 1.23$
fm$^{-2}$. Experimental values are estimates taken from the Review of Particle
Data~\cite{[Barnett]}. 

\medskip

\begin{tabular}{llrrrrr} \hline
\multicolumn{1}{c}{}
&\multicolumn{1}{c}{} 
&\multicolumn{1}{c}{Exp} 
&\multicolumn{1}{c}{(0s)$^3$} 
&\multicolumn{1}{c}{KI} 
&\multicolumn{1}{c}{G1} 
&\multicolumn{1}{c}{G2} 
 \\
\hline
P$_{33}(1232)$ &$A_{1/2}$ &$-140\pm 5\ $  &$-101$ & $-94$ & $-86$ & $-75$ \\
               &$A_{3/2}$ &$-258\pm 6\ $  &$-174$ &$-170$ &$-149$ &$-131$ \\
\hline
D$_{13}(1520)$ &$A^p_{1/2}$ &$-24\pm 9\ $   & $-22$ & $-22$ &  $-7$ &  $-71$ \\
               &$A^n_{1/2}$ &$-59\pm 9\ $   & $-41$ & $-39$ & $-45$ & $10$ \\
               &$A^p_{3/2}$ &$166\pm 5\ $   & $125$ & $124$ & $136$ & $43$ \\
               &$A^n_{3/2}$ &$-139\pm 11$   &$-125$ &$-124$ & $-91$ & $-24$ \\
\hline
F$_{15}(1680)$ &$A^p_{1/2}$ &$-15\pm 6\ $ &$\sim 0$ &   $1$ &   $7$ & $-57$ \\
               &$A^n_{1/2}$ & $29\pm 10$    &  $34$ &  $21$ &  $18$ &  $57$ \\
               &$A^p_{3/2}$ &$133\pm 12$    &  $73$ &  $89$ &  $77$ &  $27$ \\
               &$A^n_{3/2}$ &$-33\pm 9\ $   &   $0$ & $-25$ & $-21$ &  $-9$ \\
\hline
\end{tabular}

\vskip 1truecm

\noindent Table 4. The helicity amplitudes for photon decay in units $10^{-3}$ 
GeV$^{-1/2}$. SS are results from ref.~\cite{[Stancu]}, LC from 
ref.~\cite{[LiClose]}, WPR from  ref.~\cite{[Warnsb]}, C from ref.~\cite{[Capstick]}, 
BIL from ref.~\cite{[Bijker]}, Hyp from ref.~\cite{[Aiello]}. Experimental values 
are estimates taken from the Review of Particle Data~\cite{[Barnett]}. 

\medskip

\leftskip -1truecm

\begin{tabular}{llrrrrrrr} \hline
\multicolumn{1}{c}{}
&\multicolumn{1}{c}{} 
&\multicolumn{1}{c}{Exp} 
&\multicolumn{1}{c}{SS}
&\multicolumn{1}{c}{LC} 
&\multicolumn{1}{c}{WPR}
&\multicolumn{1}{c}{C}
&\multicolumn{1}{c}{BIL}
&\multicolumn{1}{c}{Hyp}
 \\
\hline
P$_{33}(1232)$ &$A_{1/2}$ &$-140\pm 5\ $  &$-101$ & $-94$ & $-81$ &$-108$ & $-91$ &$-93$\\
               &$A_{3/2}$ &$-258\pm 6\ $  &$-181$ &$-162$ &$-170$ &$-186$ &$-157$ &$-162$\\
\hline
D$_{13}(1520)$ &$A^p_{1/2}$ &$-24\pm 9\ $   &  $45$ & $-47$ &  $-7$ & $-15$ & $-43$ &$-47$\\
               &$A^n_{1/2}$ &$-59\pm 9\ $   & $-88$ & $-75$ & $-36$ & $-38$ & $-27$ &$-14$\\
               &$A^p_{3/2}$ &$166\pm 5\ $   & $202$ & $117$ &  $63$ & $134$ & $109$ &$85$\\
               &$A^n_{3/2}$ &$-139\pm 11$   &$-201$ &$-127$ &$-118$ &$-114$ &$-109$ &$-79$\\
\hline
F$_{15}(1680)$ &$A^p_{1/2}$ &$-15\pm 6\ $   &  $22$ &  $-8$ &  $18$ & $-38$ &  $-4$ &$-29$\\
               &$A^n_{1/2}$ & $29\pm 10$    &   $2$ &  $11$ &  $-4$ &  $19$ &  $40$ &$37$\\
               &$A^p_{3/2}$ &$133\pm 12$    &  $94$ & $105$ &  $87$ &  $56$ &  $80$ &$32$\\
               &$A^n_{3/2}$ &$-33\pm 9\ $   & $-30$ & $-43$ & $-18$ & $-23$ &   $0$ &$14$\\
\hline
\end{tabular}

\leftskip 0pt

\vskip 1truecm

\noindent Table 5. The total ($\sigma_T$) and backward 
($d\sigma/d\Omega\vert_{\theta=\pi}$) cross section for $\pi^0$ photoproduction 
on the proton (p) and neutron (n). Notations as in Table 3. Data as reported in 
ref.~\cite{[Copley]}.

\medskip

\leftskip -1truecm

\begin{tabular}{llrrrrrrrrr} \hline
\multicolumn{1}{c}{}
&\multicolumn{1}{c}{} 
&\multicolumn{1}{c}{} 
&\multicolumn{1}{c}{Exp} 
&\multicolumn{1}{c}{(0s)$^3$} 
&\multicolumn{1}{c}{KI} 
&\multicolumn{1}{c}{G1} 
&\multicolumn{1}{c}{G2} 
 \\
\hline
P$_{33}(1232)$  & $\sigma_T$ ($\mu$b) &                        
                 &$241.0$           &$133.8$    &$125.1$   &$97.6$  &$75.3$\\
                & $d\sigma/d\Omega\vert_{\theta=\pi}$ ($\mu$b sr$^{-1}$) &
                 &$7.0$             &$5.3$      &$4.6$     &$3.8$   &$3.0$\\
\hline
D$_{13}(1520)$  & $\sigma_T$ ($\mu$b) &p                          
                 &$16.6$            &$11.2$     &$11.1$    &$13.0$  &$4.9$\\
                &                     &n                       
                 &$12.1$            &$12.0$     &$11.8$    &$7.2$   &$0.5$\\
                &$d\sigma/d\Omega\vert_{\theta=\pi}$ ($\mu$b sr$^{-1}$) &p
                 &$0.08$            &$0.06$     &$0.05$    &$0.005$ &$0.56$\\
                &                                                       &n 
                 &                  &$0.18$     &$0.17$    &$0.22$  &$0.01$\\
\hline
F$_{15}(1680)$  & $\sigma_T$ ($\mu$b) &p                      
                 &$13.3$            &$3.7$      &$5.6$     &$4.1$   &$2.8$\\
                &                     &n                       
                 &$9.3$             &$0.8$      &$0.8$     &$0.5$   &$2.3$\\
                & $d\sigma/d\Omega\vert_{\theta=\pi}$ ($\mu$b sr$^{-1}$) &p
                 &$0$               &$<10^{-4}$ &$10^{-4}$ &$0.01$  &$0.54$\\
                &                                                       &n 
                 &                  &$0.19$     &$0.08$    &$0.05$  &$0.53$\\
\hline
\end{tabular}

\leftskip 0pt

\vskip 2truecm

\noindent Table 6. The helicity amplitudes for photon decay of the 
P$_{33}(1232)$ resonance in units $10^{-3}$ GeV$^{-1/2}$. R is the result of
ref.~\cite{[Robson]}; WW, B1 and B2 are calculated with the configuration
mixing and model parameters indicated by the corresponding labels in
Tables 1 and 2.
Experimental values are  estimates taken from the Review of Particle
Data~\cite{[Barnett]}. 

\medskip

\leftskip 2truecm

\begin{tabular}{lrrrrr} \hline
\multicolumn{1}{c}{}
&\multicolumn{1}{c}{Exp} 
&\multicolumn{1}{c}{R} 
&\multicolumn{1}{c}{WW} 
&\multicolumn{1}{c}{B1} 
&\multicolumn{1}{c}{B2} 
 \\
\hline
$A_{1/2}$ &$-140\pm 5\ $  & $-69$ & $-69$ & $-76$ &$-116$ \\
$A_{3/2}$ &$-258\pm 6\ $  &$-120$ &$-116$ &$-130$ &$-203$ \\
\hline
\end{tabular}

\clearpage

\leftskip 0pt

\noindent Table 7. The helicity amplitudes for photon decay of the P$_{33}(1232)$
resonance in units $10^{-3}$ GeV$^{-1/2}$. $\pi$ (0s)$^3$ and no $\pi$ (0s)$^3$ 
without configuration mixing and with the parameters quoted in Table 2; B1 and B2 
with complete configuration mixing and parameters as quoted in Tables 1 and 2.
Experimental values are estimates taken from the Review of  Particle
Data~\cite{[Barnett]}. 

\medskip

\begin{tabular}{llrrrr} \hline
\multicolumn{1}{c}{}
&\multicolumn{1}{c}{Contributions} 
&\multicolumn{1}{c}{$\pi$ (0s)$^3$} 
&\multicolumn{1}{c}{no $\pi$ (0s)$^3$} 
&\multicolumn{1}{c}{B1} 
&\multicolumn{1}{c}{B2} 
 \\
\hline
$A_{1/2}$ &&&&& \\ 
&1-body     &$-104.0$  &$-104.4$ &$-106.3$ &$-106.7$   \\
&gluon exch. & $-10.5$  & $-15.6$ &  $-3.7$ &  $-9.0$   \\
&pion exch.  &  $27.2$  &         &  $33.6$ &           \\
&total      & $-87.3$  &$-120.0$ & $-76.4$ &$-115.7$   \\
\hline
$A_{3/2}$ &&&&& \\
&1-body     &$-180.2$  &$-180.8$ &$-185.5$ &$-186.7$   \\
&gluon exch. & $-19.2$  & $-28.5$ &  $-6.9$ & $-16.5$   \\
&pion exch.  &  $47.2$  &         &  $61.9$ &           \\
&total      &$-152.2$  &$-209.3$ &$-130.5$ &$-203.2$   \\
\hline
\end{tabular}

\clearpage

\noindent Table 8. The helicity amplitudes for photon decay of the
D$_{13}(1520)$ and F$_{15}(1680)$ resonances in units $10^{-3}$
GeV$^{-1/2}$. The model parameters B1 and B2 are taken from Table 2  without
configuration mixing, labelled (0s)$^3$, and with the configuration mixings G,
B1 and B2 from Table 1. Experimental values are estimates taken from the Review
of Particle Data~\cite{[Barnett]}. 

\medskip

\leftskip -1truecm

\begin{tabular}{lrrrrrrr} \hline
\multicolumn{1}{c}{}
&\multicolumn{1}{c}{Exp} 
&\multicolumn{1}{c}{B1--(0s)$^3$}
&\multicolumn{1}{c}{B2--(0s)$^3$} 
&\multicolumn{1}{c}{B1--G}
&\multicolumn{1}{c}{B2--G}
&\multicolumn{1}{c}{B1--B1}
&\multicolumn{1}{c}{B2--B2}
\\
\hline
D$_{13}(1520)$ &&&&&&& \\
$A^p_{1/2}$ &$-24\pm 9\ $   &  $19$ &  $-1$ &  $16$ &  $-3$ &   $4$ & $-9$ \\
$A^n_{1/2}$ &$-59\pm 9\ $   & $-64$ & $-41$ & $-60$ & $-38$ & $-42$ &$-27$ \\
$A^p_{3/2}$ &$166\pm 5\ $   & $158$ & $109$ & $151$ & $104$ & $118$ &$82$  \\
$A^n_{3/2}$ &$-139\pm 11$   & $-77$ & $-25$ & $-71$ & $-22$ & $-49$ &$-10$ \\
\hline
F$_{15}(1680)$ &&&&&&& \\
$A^p_{1/2}$ &$-15\pm 6\ $   &  $13$ &  $20$ &   $9$ &  $21$ &  $11$ &$20$  \\
$A^n_{1/2}$ & $29\pm 10$    &  $23$ &  $20$ &   $7$ &  $10$ &  $14$ &$8$   \\
$A^p_{3/2}$ &$133\pm 12$    &  $34$ &  $51$ &  $45$ &  $69$ &  $46$ &$69$  \\
$A^n_{3/2}$ &$-33\pm 9\ $   &  $10$ &  $-1$ & $-20$ & $-24$ & $-11$ &$-27$ \\
\hline
\end{tabular}

\leftskip 0pt

\vskip 2truecm

\noindent Table 9. The total ($\sigma_T$) and backward 
($d\sigma/d\Omega\vert_{\theta=\pi}$) cross section for $\pi^0$ photoproduction 
on the proton (p) and neutron (n). Notations as in Table 7. Data as reported 
in ref.~\cite{[Copley]}.

\medskip

\leftskip -1truecm

\begin{tabular}{llrrrrrrrrr} \hline
\multicolumn{1}{c}{}
&\multicolumn{1}{c}{} 
&\multicolumn{1}{c}{} 
&\multicolumn{1}{c}{Exp} 
&\multicolumn{1}{c}{$\pi$ (0s)$^3$} 
&\multicolumn{1}{c}{no $\pi$ (0s)$^3$} 
&\multicolumn{1}{c}{B1} 
&\multicolumn{1}{c}{B2} 
 \\
\hline
P$_{33}(1232)$  & $\sigma_T$ ($\mu$b) &                        
                 &$241.0$           &$101.6$    &$192.1$   &$75.4$  &$180.5$\\
                & $d\sigma/d\Omega\vert_{\theta=\pi}$ ($\mu$b sr$^{-1}$) &
                 &$7.0$             &$4.0$      &$7.6$     &$3.0$   &$7.0$\\
\hline
D$_{13}(1520)$  & $\sigma_T$ ($\mu$b) &p                          
                 &$16.6$            & $8.7$     & $4.5$    &$16.0$  &$7.6$\\
                &                     &n                       
                 &$12.1$            & $8.6$     & $2.9$    &$6.0$   &$1.3$\\
                &$d\sigma/d\Omega\vert_{\theta=\pi}$ ($\mu$b sr$^{-1}$) &p
                 &$0.08$            &$0.13$      &$0.28$    &$0.03$  &$10^{-3}$\\
                &                                                       &n 
                 &                  &$0.012$     &$0.02$    &$0.40$  &$0.16$\\
\hline
F$_{15}(1680)$  & $\sigma_T$ ($\mu$b) &p                      
                 &$13.3$            &$3.3$      &$4.8$     &$1.5$   &$3.6$\\
                &                     &n                       
                 &$9.3$             &$1.7$      &$1.6$     &$0.3$   &$0.5$\\
                & $d\sigma/d\Omega\vert_{\theta=\pi}$ ($\mu$b sr$^{-1}$) &p
                 &$0$               &$0.13$     &$0.12$    &$0.01$  &$0.08$\\
                &                                                       &n 
                 &                  &$0.40$     &$0.39$    &$0.06$  &$0.02$\\
\hline
\end{tabular}

\leftskip 0pt

\end{document}